\documentclass[prb,twocolumn,groupedaddress]{revtex4b4}
\usepackage{graphicx}
\begin{document}
\title{Modelling the structure of clusters of C$_{60}$ molecules}
\author{Jonathan P.~K.~Doye}
\email{jon@clust.ch.cam.ac.uk}
\affiliation{University Chemical Laboratory, Lensfield Road, Cambridge CB2 1EW, United Kingdom}
\author{David J.~Wales}
\affiliation{University Chemical Laboratory, Lensfield Road, Cambridge CB2 1EW, United Kingdom}
\author{Wolfgang Branz}
\affiliation{Max-Planck-Institut f\"{u}r Festk\"{o}rperforschung, Heisenbergerstr.\ 1,
70569 Stuttgart, Germany}
\author{Florent Calvo}
\affiliation{Laboratoire de Physique Quantique,
IRSAMC, Universit\'e Paul Sabatier, 118 Route de Narbonne, F31062 Toulouse Cedex.} 
\date{\today}
\begin{abstract}
We locate putative global minima for (C$_{60}$)$_N$ clusters 
modelled by the potential of Pacheco and Prates-Ramalho up to $N$=105. 
These minima are based on icosahedral packing up to $N$=15, 
but above this size the lowest-energy structures are decahedral or close-packed.
Although structures based on the 98-molecule Leary tetrahedron, which have been 
inferred from experiment, are not lowest in energy for this potential, an examination
of the energetics of a growth sequence leading to the Leary tetrahedron lends
further support to the experimental assignments.
An analysis of the potential energy surface topography and the thermodynamics of
two example clusters indicates that the multiple-funnel topography is likely to
have a strong influence on the dynamics of structure formation and that
solid-solid transitions driven by differences in vibrational entropy are likely to be common.
\end{abstract}
\maketitle

\section{Introduction}
\label{sect:intro}
There has been much interest in the condensed-phase properties of C$_{60}$
molecules because of their unusual intermolecular potential.
At high temperature the molecules can rotate freely and so they act
as large `pseudo-atoms' with interactions that are extremely 
short-ranged relative to the large equilibrium pair separation.
Consequently the properties of C$_{60}$ can extend beyond the 
range of behaviour observed for atomic materials. 
For example, there are theoretical predictions that 
the liquid phase is unstable\cite{Hagen93,Broughton97} or 
marginally stable\cite{ACheng,Caccamo97,Hasegawa99,Ferreira00} 
with the precise results depending on the potential (and somewhat on the 
methodology\cite{Hasegawa00}) used.

There has also been considerable interest in the structural properties
of clusters of C$_{60}$ molecules, 
prompted by the first experiments of Martin {\it et al.\/} on 
positively-charged clusters.\cite{Martin93}
This mass spectroscopic study revealed magic numbers
for clusters with less than 150 molecules 
that are indicative of structures based upon Mackay 
icosahedra.\cite{Mackay} The stability of the icosahedral (C$_{60}$)$_{13}$
has since been further illustrated by Hansen {\it et al.}.\cite{Hansen96,Hansen97}
However, subsequent calculations using 
the spherically-averaged Girifalco potential \cite{Girifalco} found that icosahedral
structures are only lowest in energy up to $N$=13.\cite{Rey94,Wales94a,Doye96d} 
Above this size the structures are either decahedral or close-packed.
The rapid emergence of bulk-like structures for this potential contrasts with
many other atomic clusters where icosahedral structures can persist up
to large sizes, for example, up to at least $20\,000$ atoms for sodium.\cite{Martin90}
This behaviour, and the marginal stability of the liquid phase, has a 
common origin in the narrowness of the C$_{60}$ intermolecular potential well.
As the range of a potential is decreased, there is an increasing energetic
penalty for strained structures, be they icosahedra\cite{Doye95c}
or liquid configurations,\cite{Doye96a,Doye96b}  
which involve nearest-neighbour distances that deviate 
from the equilibrium pair distance.

One possible cause of the discrepancy between experiment 
and theoretical calculations is the isotropic nature of the Girifalco potential.
However, calculations using an all-atom model also found that
icosahedra are only lowest in energy at small 
sizes, albeit to slightly larger sizes ($N$=16) than 
for the Girifalco potential.\cite{Rey97,Doye97c,Garcia97}

This situation led Shvartsburg and Jarrold to investigate the possibility 
of distinguishing icosahedral from decahedral and close-packed
isomers by their mobility. However, the differences in the computed mobilities 
are small and were comparable to the experimental resolution.\cite{Shvartsburg96}
More recently, Branz {\it et al.\/} performed some new mass spectroscopic 
experiments on (C$_{60}$)$_N$ clusters to try to obtain further
insights into the difference between theory and experiment.\cite{Branz00}
They found that the observed structures are independent of the sign and
magnitude of the charge. However, temperature was found to be a key variable 
in determining the structure. 
The initial cold as-grown clusters showed no magic numbers. 
Only on annealing at higher temperatures are the magic numbers revealed,
as the relative evaporation rates cause larger populations to develop
in those clusters that are more resistant to evaporation.
After annealing at 490$\,$K, as in the previous experiment, magic numbers 
consistent with icosahedral clusters are obtained. However,
annealing at 585$\,$K reveals a new set of magic numbers that correspond to 
non-icosahedral clusters.\cite{timescale} As well as sizes associated with 
face-centred-cubic (fcc) and decahedral packing, particularly prominent
were magic numbers associated with the recently discovered Leary tetrahedron.\cite{Leary99}

There are two possible explanations for this temperature dependence. 
Firstly, the results could reflect changes in the thermodynamically most
stable structures with temperature. Such solid-solid transitions have now
been observed in a variety of 
systems.\cite{Doye95c,Doye98a,Cleveland98,Cleveland99,Doye99c,Neirotti00,Doye01b,Calvo01d}  
However, for all these examples icosahedral structures are more stable at higher temperature,
and are favoured because they have a larger vibrational entropy.\cite{Doye01b}
Furthermore, the theoretical calculations for (C$_{60}$)$_N$ suggest that 
the non-icosahedral structures are lowest in energy and so would be favoured at
low temperature. Indeed, such a transition to a high-temperature icosahedral
structure has been located for (C$_{60}$)$_{14}$ interacting with 
the Girifalco potential\cite{Calvo01d} and has also been suggested for some
larger clusters.\cite{Zhang00} As the behaviour of these solid-solid transitions
is opposite to that of the experiment, this explanation is unlikely.

Secondly, the icosahedral structures could be more accessible during the 
growth and annealing of the clusters, with escape from the icosahedral region of
configuration space into the state with lowest free energy only being possible at 
the highest temperatures. This interpretation has support from a variety of sources.
For example, it has been shown that it is possible, under certain conditions,
to preferentially grow metastable icosahedral clusters for silver.\cite{Baletto01}
Furthermore, analyses of the potential energy surface (PES) topography of those Lennard-Jones
clusters with non-icosahedral global minima has shown that large energy barriers exist 
for interconversion of the icosahedral isomers and the global minimum, and that 
the icosahedral funnel is much wider.\cite{Doye99f} 
Thus, on relaxation down the PES, the system
is likely to become trapped in the icosahedral region of configuration space. 
This trapping is partly due to the greater structural similarity to the polytetrahedral 
configurations typical of the liquid, but also to the general energetic preference
that Lennard-Jones clusters have for icosahedral geometries.\cite{Northby87} 
The latter would not hold for clusters of C$_{60}$ molecules. 
Finally, short-ranged potentials have been 
shown to lead to relatively large barriers\cite{Wales94b,Miller99a}
and consequently to much slower dynamics.\cite{Miller99b}
Therefore, it is plausible that kinetic effects could dominate for (C$_{60}$)$_N$ clusters
on the experimental time scales, even for the relatively small clusters studied. 

Many of the studies of the condensed-phase properties of C$_{60}$ have used the Girifalco 
potential.\cite{Hagen93,ACheng,Caccamo97,Hasegawa99,Rey94,Wales94a,Doye96d,Gallego99}
However, a new single-site potential has recently been derived by Pacheco 
and Prates-Ramalho (PPR) that gives an improved description of many properties of 
bulk C$_{60}$, including, for example, the dependence of the density of the crystal
on pressure.\cite{Pacheco97} 
Furthermore, the high-temperature thermodynamics of clusters interacting with this
potential are in reasonable agreement with those of an all-atom potential.\cite{Calvo01c}
Here, we see if this potential can help to explain the experimental observations
of Branz {\it et al.\/}.\cite{Branz00}
Global optimization has been previously attempted for the PPR potential.\cite{Zhang00,Luo99} 
However, the conclusions of these papers are contradictory and, 
as we shall see, many of the putative global minima are sub-optimal.
Here, as well as locating putative global minima for this potential, we also
consider the most stable growth sequence for structures based on the 
experimentally observed Leary tetrahedra,
and examine the PES topography and the thermodynamics of
two example clusters. 

\section{Methods}
\label{sect:methods}

The PPR potential consists of a pair potential plus a three-body term of the standard
Axilrod-Teller form.\cite{Axilrod} The potential energy of the cluster is therefore given by
\begin{equation}
E=\sum_{i<j} V_{\rm pair}(r_{ij}) + C_{\rm AT} \sum_{i<j<k} V_{\rm AT}(r_{ij},r_{ik},r_{jk}),
\end{equation}
where $C_{\rm AT}$ gives the magnitude of the Axilrod-Teller term.
The pair potential consists of a Morse form for the short-range repulsion, 
a van der Waals expansion for the long-range attraction and a Fermi function
to describe the crossover between these two regimes. 
The forms of these functions and the associated parameters are 
given in Ref.\ \onlinecite{Pacheco97}.

\begin{figure}
\begin{center}
\includegraphics[width=8.2cm]{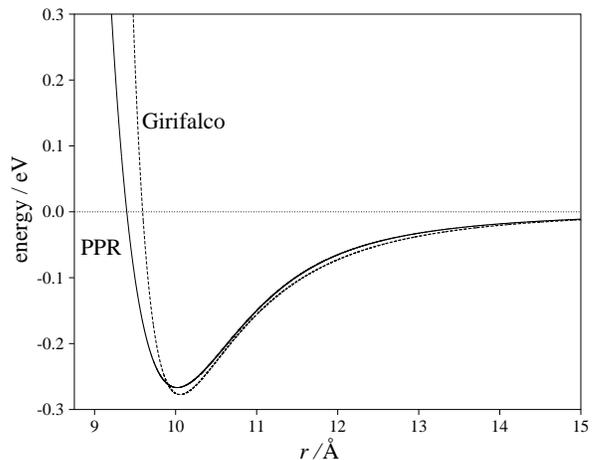}
\caption{\label{fig:potential} Comparison of the PPR and Girifalco pair potentials.}
\end{center}
\end{figure}

As one can see from Figure \ref{fig:potential}, the PPR pair potential has a softer repulsion 
than the Girifalco potential, leading to a wider potential well. 
This effect can be quantified by matching the second derivative at the
bottom of the well to that of the Morse potential, 
\begin{equation}
V_M(r)=\epsilon\exp\left[\rho(1-r/r_{\rm eq})\right]
       \left(\exp\left[\rho(1-r/r_{\rm eq})\right]-2 \right),
\end{equation}
where $\epsilon$ is the pair well depth and $r_{\rm eq}$ is the equilibrium pair distance.
The Morse potential becomes increasingly narrow as the parameter $\rho$ increases.
The curvature at the bottom of the Morse well is $2\rho^2$ 
when the units of energy and distance are the pair well depth and equilibrium pair distance.
This result can then be used to obtain a value of 
$\rho_{\rm eff}$, a measure of the effective range, for each
potential. This analysis has been done previously for the Girifalco potential: 
$\rho_{\rm eff}^{\rm G}$=13.62.\cite{WalesU94} 
For the PPR potential, however, $\rho_{\rm eff}^{\rm PPR}$=11.28.
This difference in $\rho_{\rm eff}$ has a well-understood effect on the resulting properties 
of the liquid\cite{Hagen94,Doye96a,Doye96b} and of clusters.\cite{Doye95c,Doye97d}
For example, it will make the bulk PPR liquid phase more stable, as has been observed.\cite{Ferreira00} 
More importantly for this study, it will make icosahedral structures more stable than
for the Girifalco potential. 

The three-body Axilrod-Teller term always gives rise to a positive contribution 
to the energy for a compact structure. For example, its inclusion leads to a 6\% increase
in the energy of the bulk C$_{60}$ crystal.\cite{Pacheco97}
However, in the global optimization study of Ref.\ \onlinecite{Luo99}, the supposed global 
minima for the full PPR potential had a lower potential energy than those when only the 
pair potential was used. This is clearly wrong and presumably must have been due to an 
error when coding the Axilrod-Teller term. Furthermore, in the other optimization study
of PPR clusters, putative global minima were only located for the pair potential, because of the
greater computational cost of the three-body term.\cite{Zhang00}

The most unfavourable common configuration for the three-body Axilrod-Teller term is three
nearest-neighbour molecules arranged in an equilateral triangle.\cite{Wales90a} 
Therefore, one would expect the three-body energy to be larger for those structures 
that have more nearest neighbours, more polytetrahedral character and 
close-packed surfaces. Thus the Axilrod-Teller term 
disfavours icosahedral structures the most.\cite{Up92}

To locate the global minima for the PPR potential we used basin-hopping\cite{WalesD97} 
(Monte Carlo plus minimization\cite{Li87a}), which has proved to be a very effective
method for a variety of cluster systems. This approach\cite{WalesD97,WalesS99} 
and the reasons for its success\cite{Doye98a,Doye98e,Doye01c} have been 
described in detail elsewhere. The only specific modification for the present application 
was to reduce the computational cost of the minimizations by only switching on the three-body 
interactions close to convergence. Such a `guiding function' approach has 
previously been suggested and exploited by Hartke.\cite{Hartke96}
As well as basin-hopping, we also reoptimized a large database of minima that we 
had obtained from previous optimization studies on Lennard-Jones,\cite{WalesD97} 
Morse\cite{Doye95c,Doye97d} and Girifalco\cite{Doye96d} clusters. 
Most of the global minima that we located were contained within this database.

To generate the samples of minima from which the disconnectivity graphs and
the thermodynamics in Section \ref{sect:3855} were calculated, 
we used the same methods as those we have previously applied to LJ\cite{Doye99f,Doye99c,Doye00c} 
and Morse\cite{Miller99a} clusters.
We thereby obtain large samples of connected minima and transition states that provide
good representations of the low-energy regions of the PES.
The approach involves repeated applications of eigenvector-following\cite{Cerjan} to
find new transition states and the minima they connect, as described in detail elsewhere.\cite{WalesDMMW00}

\section{Global minima}
\label{sect:gmin}

Putative global minima were located for the full PPR potential up to $N$=105, 
the size range of interest for comparison with the high temperature experiments 
of Branz {\it et al.\/}.\cite{Branz00}
The energies and point groups of these structures are given in Table \ref{table:gmin}
and coordinates are available on the world-wide web from the Cambridge Cluster Database.\cite{Web}

As expected, the energies of the global minima differ from those reported in 
Ref.\ \onlinecite{Luo99}, because of the error in the Axilrod-Teller term. 
To facilitate comparisons with the other results in Refs.\ \onlinecite{Zhang00} and 
\onlinecite{Luo99}, we also report the energies of these global minima when 
reoptimized for the PPR pair potential (Table \ref{table:gmin}).
The resulting energies will not necessarily be the global optima for the pair potential, but
they should provide a good upper bound. Comparisons with these energies show that the 
putative global minima for the PPR pair potential given in Ref.\ \onlinecite{Luo99}
are sub-optimal for $N\ge 19$ and those in Ref.\ \onlinecite{Zhang00} are sub-optimal
for $N$=18, 26, 33--36 and $N\ge 40$. Only for $N$=16 and 29 are lower two-body energies
obtained in these papers,\cite{Zhang00,Luo99} indicating that for these cluster sizes
the introduction of the Axilrod-Teller term leads to a change in the structure of the
global minimum. For $N$=16 the relevant minimum is icosahedral with $E$=$-13.006550\,$eV
and $E_2$=$-13.447202\,$eV, and for $N$=29 it is an alternative decahedral 
minimum with $E$=$-27.540094\,$eV and $E_2$=$-28.516282\,$eV.

\begin{figure}
\begin{center}
\includegraphics[width=8.4cm]{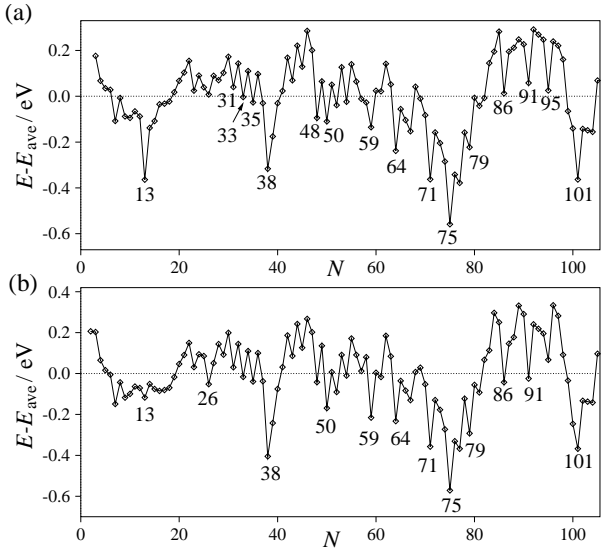}
\caption{\label{fig:E}Energies of the global minima for (C$_{60}$)$_N$ clusters modelled 
by the (a) PPR and (b) Girifalco potentials. The energy zero is taken to be $E_{\rm ave}$, 
a four-parameter fit to the energies of the global minima. 
For the PPR potential $E_{\rm ave}$/eV$=-1.6537 N + 2.1901 N^{2/3} + 0.1263 N^{1/3} -0.7467 $.
For the Girifalco potential $E_{\rm ave}$/eV$=-1.789 N + 2.294 N^{2/3} + 0.5907 N^{1/3} -1.292$.
}
\end{center}
\end{figure}

\begin{figure*}
\begin{center}
\includegraphics[width=18cm]{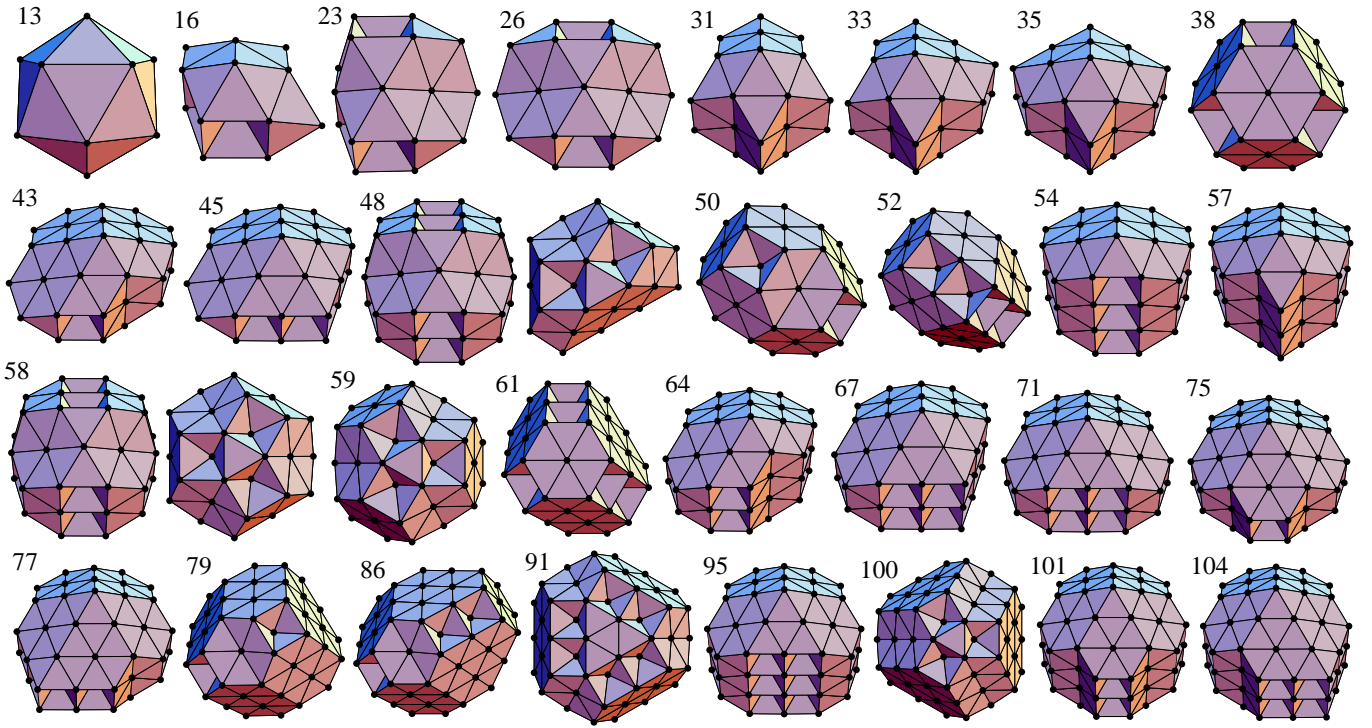}
\caption{\label{fig:structures} A selection of the global minima for the PPR potential. 
Each molecule is represented by a point at the molecular centre. 
The two views of the 48- and 58-molecule
clusters illustrate the relationship of these structures to both decahedra and the Leary 
tetrahedron.}
\end{center}
\end{figure*}

In Figure \ref{fig:E}a we plot the energies of the PPR global minima so that 
particularly stable clusters stand out. We also provide a comparison with results
for the Girifalco potential (Figure \ref{fig:E}b), partly because a significant number of the 
putative global minima in Ref.\ \onlinecite{Doye96d} have now been improved 
(an up-to-date list is maintained at the Cambridge Cluster Database\cite{Web}).
A selection of the global minima are depicted in Figure \ref{fig:structures}.

At $N$=13 the icosahedral global minimum is noticeably more stable than for the Girifalco
potential, which is consistent with the PPR potential's slightly wider well. 
Furthermore, icosahedral structures are lowest in energy up to 15 molecules, or up to 
16 molecules if the Axilrod-Teller term in the potential is not included---as noted in Section 
\ref{sect:methods} this term slightly disfavours icosahedral structures. 
However, contrary to the claims of Refs.\ \onlinecite{Zhang00} and 
\onlinecite{Luo99} there are no further icosahedral global minimum above this size. 
Instead decahedral or close-packed clusters are lowest in energy. 

To illustrate how the relative stabilities of icosahedral structures with respect to 
the global minimum evolve with size, we give the energy differences for examples
where particularly stable fcc and icosahedral forms are available.
At $N$=38 the difference in energy between the fcc truncated octahedron and the lowest-energy
icosahedral structure is $0.582\,$eV. 
However, at $N$=55 the Mackay icosahedron is $0.297\,$eV above the decahedral global minimum.
When the Axilrod-Teller interactions are not included this difference decreases to $0.159\,$eV, again 
illustrating that this term slightly disfavours icosahedra. 
In contrast, the energy difference for the Girifalco potential is a considerably larger $1.997\,$eV.
These results are consistent with the behaviour expected from the respective values of $\rho_{\rm eff}$:
the Mackay icosahedron is lowest in energy for the Morse potential up to $\rho$=11.15 
and then becomes increasingly unstable as the width of the potential decreases further.

\begin{figure*}
\begin{center}
\includegraphics[width=18cm]{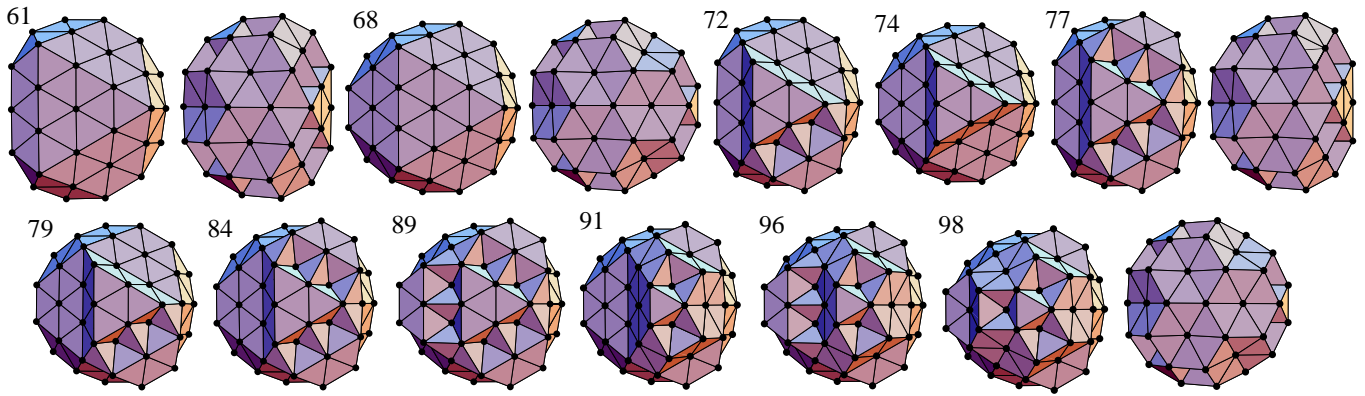}
\caption{\label{fig:Leary} Particularly stable clusters on the growth sequence
leading to the Leary tetrahedron. When two viewpoints are given they correspond to the front
and back of the cluster.}
\end{center}
\end{figure*}

The pronounced minima in Figure \ref{fig:E} correspond to particularly low-energy
structures that could potentially give rise to magic numbers.
Comparing (a) and (b) of Figure \ref{fig:E} it is apparent that the patterns
are very similar, indicating that the two C$_{60}$ potentials favour similar structures. Prominent
in both figures are the same particularly stable decahedral and close-packed forms, most of which
are illustrated in Figure \ref{fig:structures}.
The main differences are that, for the Girifalco potential, the feature due to
(C$_{60}$)$_{13}$ is much less pronounced, for the reasons explained above,
and that the close-packed structures are somewhat more favoured. The latter feature is
again due to the shorter range of the Girifalco potential, as is the greater number of
close-packed global minima for the Girifalco potential. 

There is some correspondence between Figure 2 and the high temperature magic numbers 
observed experimentally.\cite{Branz00} 
Small peaks in the mass spectrum at $N$=31, 33 and 35 probably 
correspond to a series of small decahedra. Similarly, the peak at $N$=38 can
be assigned to the fcc truncated octahedron, and the peaks at $N$=64, 71 and 75 
are probably due to Marks decahedra. Particularly interesting is 
the feature in Figure \ref{fig:E} corresponding to the decahedral (C$_{60}$)$_{48}$
global minimum, because the most prominent peak in the mass spectrum occurs at $N$=48. 
Like a number of the other decahedral global minima, this structure has
an overlayer on the $\{111\}$ faces in sites which are hcp with respect 
to the five slightly strained fcc tetrahedra that are the basis of decahedral structures.
In fact, many of these structures are fragments of larger Mackay icosahedra
with the apex of the decahedron corresponding to what would become the centre of the icosahedron. 
Indeed, growth simulations for silver have indicated that the addition of this overlayer
does provide a natural pathway to the growth of larger icosahedra.\cite{Baletto01}

However, many of the particularly stable structures, especially for $N$$\ge$50, do not
correspond to magic numbers in the experiment.
There are no experimental peaks corresponding to the close-packed tetrahedra
at $N$=59 and 100, nor to the twinned truncated octahedra at $N$=50 and 79 and the 101-molecule
Marks decahedron.
Instead, peaks occur in the mass spectrum at $N$=58, 61, 68, 77, 84, 91, 96 and 98, which cannot
be simply explained by reference to the global minima of model C$_{60}$ potentials.
These peaks have previously been assigned\cite{Branz00} to structures based on a 98-molecule Leary 
tetrahedron.\cite{Leary99}
This recently discovered structure is illustrated in Figure \ref{fig:Leary}. It can be described
in terms of five fcc tetrahedra arranged into a stellated tetrahedron with apex molecules removed
and 7-molecule hexagonal patches covering the edges of the central tetrahedron. That this 
structure is not the global minimum for the PPR or Girifalco potential is consistent with 
the values of $\rho_{\rm eff}$. For the Morse potential it is only the global minimum
for $6.91<\rho<9.45$ because it has a strain energy intermediate between decahedra and icosahedra.

\begin{figure}
\begin{center}
\includegraphics[width=8.2cm]{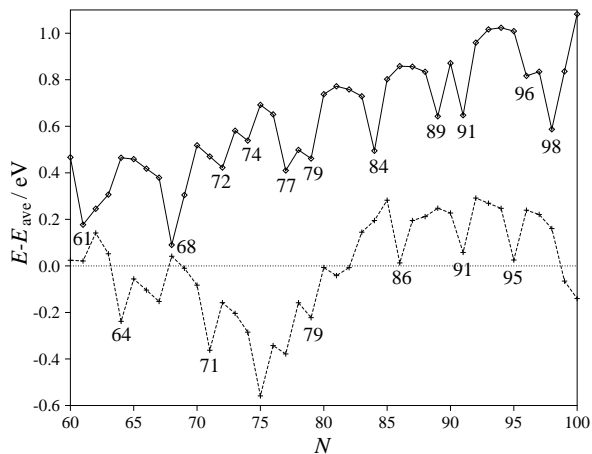}
\caption{\label{fig:Eleary} Energies of structures based upon the 98-molecule Leary
tetrahedron (solid line) compared to the energies of the global minima (dashed line) for
the PPR potential.}
\end{center}
\end{figure}

To attempt to add further weight to the experimental assignments we calculated the energies 
of a series of structures derived from the Leary tetrahedron. Although they are not 
lowest in energy, the size variation of their energies, plotted for $60<N<100$ in 
Figure \ref{fig:Eleary}, is insightful. There is remarkable agreement between
the features in this graph and the remaining magic numbers. 

The stable structures for the Leary tetrahedral growth sequence are illustrated in 
Figure \ref{fig:Leary}. The 91-molecule structure is derived by removing a hexagonal 
patch from over one of the edges of the central tetrahedron. 
In contrast, to obtain the 89-molecule structure one of the points of the Leary 
tetrahedron is further truncated. By a similar truncation the (C$_{60}$)$_{91}$ global
minimum can be derived from the 100-molecule tetrahedral global minimum (Figure \ref{fig:structures}).
Most of the other structures can be obtained by a combination of these two changes. The
84-molecule structure is derived by a truncation and the removal of an adjacent patch. 
Similarly, truncation of two of the fcc tetrahedra and the removal of the common patch gives
the 77-molecule structure. The 68-molecule structure is derived by completely removing one of the 
fcc tetrahedra and adjacent patches, thus giving a structure that is also a fragment of 
the 147-molecule Mackay icosahedron. 
The truncation of a further tetrahedron leads to the 61-molecule structure.

Interestingly, the latter two structures are global minima for Morse clusters, but 
were not located in the most recent optimization study.\cite{Doye97d}
They have energies of $E_{61}(\rho$=10)=$-252.488332\,\epsilon$ and 
$E_{68}(\rho$=10)=$-286.643320\,\epsilon$, and are lowest in energy
in the ranges $9.42<\rho<10.34$ and $7.40<\rho<11.55$, respectively.

The prominence of the peak in the mass spectrum at $N$=48 also fits with this 
preference for structures based upon the Leary tetrahedron. 
The second view of the decahedral (C$_{60}$)$_{48}$ global minimum in Figure \ref{fig:structures}
shows that it is also a fragment of the Leary tetrahedron. 
By adding an identical overlayer to the bottom $\{111\}$ faces 
of this cluster the decahedral (C$_{60}$)$_{58}$ global minimum is obtained, 
which is again a fragment of the Leary tetrahedron. 
Although this cluster is not especially stable compared to nearby sizes for the PPR
potential, its relationship to the Leary tetrahedron provides a basis for confidently assigning the remaining
magic number at $N$=58 to this structure.

\begin{figure*}
\begin{center}
\includegraphics[width=18cm]{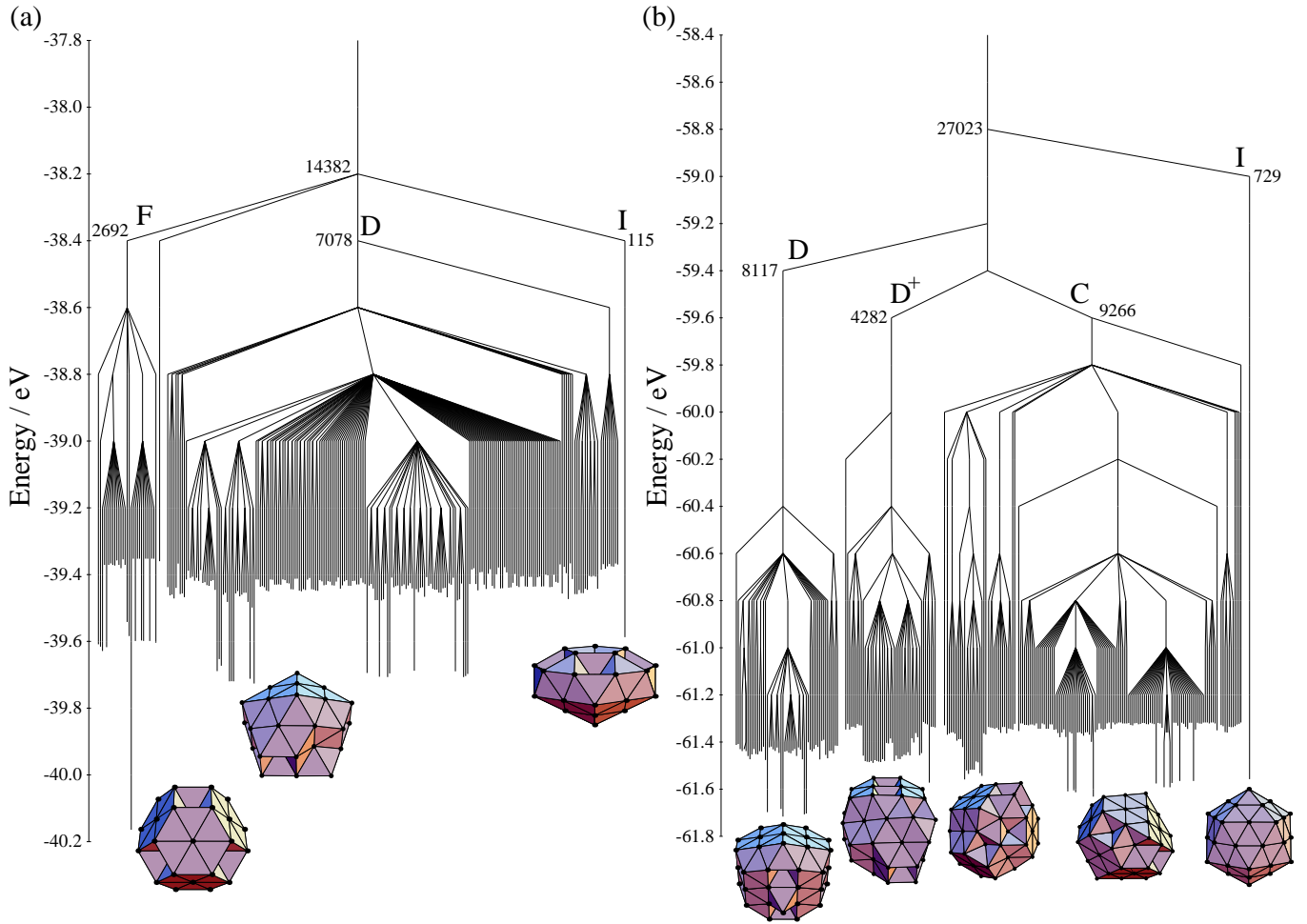}
\caption{\label{fig:tree} Disconnectivity graphs for (a) (C$_{60}$)$_{38}$ and (b) (C$_{60}$)$_{55}$. 
Lines corresponding to the lowest 250 minima are represented. Structural labels have been 
placed adjacent to the lines corresponding to the different funnels
on the PES, where, as in Table \ref{table:gmin}, I stands for icosahedral, D for decahedral 
(D$^+$ have an overlayer on the $\{111\}$ faces), F for fcc and C for close-packed. 
The numbers refer to the number of minima 
below the corresponding node in the full graph.
Pictures of the lowest-energy minima for the different structural types have been placed adjacent 
to the corresponding lines.
}
\end{center}
\end{figure*}

\section{(C$_{60}$)$_{38}$ and (C$_{60}$)$_{55}$}
\label{sect:3855}

Although the location of the global minima should be the first step in the theoretical determination
of a cluster's structure, the thermodynamics\cite{Doye01b} and dynamics\cite{Baletto00} 
can also play an important role. Here, we examine the behaviour of two example clusters in more detail. 
We choose (C$_{60}$)$_{38}$ and (C$_{60}$)$_{55}$ because particularly stable fcc (38) and 
icosahedral (55) structures are possible. Furthermore, these sizes have been extensively studied 
for the longer-ranged Lennard-Jones potential.\cite{StillD90a,Kunz93,Doye98a,Doye99c,Doye99f,Miller99b}

In order to construct disconnectivity graphs for these two clusters we have generated large samples
of minima and transition states, plus the pathways connecting them. 
As this is a computationally demanding task only the two-body component of the PPR potential was used.
The graphs for the full potential would be very similar; the main effect of the introduction of the three-body term
would be to displace the icosahedral regions of configuration space further up in energy.
We located $38\,558$ minima and $39\,959$ transition states for $N$=38 and 
$39\,043$ minima and $39\,845$ transition states for $N$=55. 

Disconnectivity graphs\cite{Becker97,WalesMW98} provide a representation of the barriers between minima
on a PES.\cite{Becker97,WalesMW98}
In a disconnectivity graph, each line ends at the energy of a minimum.
At a series of equally-spaced energy levels we determine which (sets of) minima are connected
by paths that never exceed that energy.
We then join up the lines in the disconnectivity graph at the energy level where
the corresponding (sets of) minima first become connected.
In a disconnectivity graph an ideal single-funnel\cite{Leopold,Bryngelson95} PES would be represented by a
single dominant stem associated with the global minimum to which the other minima directly join. 
For a multiple-funnel PES there would be a number of major stems that only join at high energy.
A single funnel PES is typically associated with efficient relaxation to the global minimum,
whereas for a multiple-funnel PES there is a separation of time scales between relaxation to a low-energy 
structure and interfunnel equilibrium.\cite{WalesDMMW00}

For (C$_{60}$)$_{38}$ and (C$_{60}$)$_{55}$ it is not possible to characterize the whole PES because of the huge numbers of
minima. Besides, even if we could perform such a characterization, any attempt at representation 
using a disconnectivity graph would just be obscured by the density of lines. Therefore,
we concentrate on the low-energy regions of the PES that are of most importance 
when considering structure, and only include the lines
leading to the lowest 250 minima.

For both clusters the PES tography has a multiple-funnel character, 
where each funnel corresponds to a different structural type (Figure \ref{fig:tree}). 
For this short-ranged potential, there are a number of competing low-energy morphologies that 
are separated by large energetic barriers.  
As the barriers between minima of the same structural type are typically much smaller, 
the graphs clearly separate the different morphologies. 

This behaviour contrasts with 
that for Lennard-Jones clusters, which (with some notable exceptions) typically have a
single-funnel topography that is dominated by icosahedral structures.\cite{Doye99f}
However, for the PPR potential the difference in energy between close-packed and decahedral
structures is small. Furthermore, there are often a number of close-packed forms possible
that have significant structural differences. For example, for (C$_{60}$)$_{55}$ the close-packed
region of configuration space divides up into structures that are based on the 50-molecule global minimum 
(the twinned truncated octahedron) and those based on the 59-molecule 
tetrahedral global minimum (Figure \ref{fig:structures}). 
Additionally, as well as a funnel leading to the conventional 
decahedral global minimum, there is a low-energy region of configuration space corresponding
to decahedral structures that have the $\{111\}$ faces partially covered. These latter 
structures can be constructed by adding molecules to the 48-molecule global minimum.
This complexity of the (C$_{60}$)$_{55}$ disconnectivity graph contrasts with that of 
(C$_{60}$)$_{38}$, for which there are only funnels associated with the three basic morphologies.

The disconnectivity graphs can tell us a lot about the dynamics of these two clusters.
On the disconnectivity graphs we give the numbers of minima in our sample 
that are associated with each funnel. 
Although, these are large underestimates of the true values because of the incompleteness
of our samples, they do provide a reasonable indication of the relative numbers 
of low-energy minima of each structural type, and hence of the widths of the funnels 
in configurational terms. It is particularly noticeable that there are far 
fewer low-energy icosahedral minima, as expected for a short-ranged potential.
By contrast, the number of decahedral and close-packed structures is of the same order.

Based on this information, on relaxation down the PES one would expect the system to be
more likely to end up in a close-packed or decahedral configuration. However, there are other
factors that can influence the dynamical behaviour. There is a vibrational, as well as a 
configurational, contribution to the width of a funnel. This term is larger for the 
icosahedral funnel because the associated minima are generally flatter and wider.
Furthermore, it may also be that the icosahedral structures lie, in some sense, closer to 
the liquid-like region of configuration space because of their greater polytetrahedral 
character.\cite{NelsonS,Doye96b}

Structure formation in clusters does not necessarily occur by relaxation from a high-energy
state, but can also occur by growth from a smaller pre-existing solid cluster.\cite{Baletto00} 
Although the width of the funnels is less important in this case, the disconnectivity graphs
can still be useful because they give an idea of the kinetic stability of the possible structures.
The large barriers between funnels suggest that, except at high temperature, trapping will occur 
within the funnel of the existing structure, even if the structure is metastable.
Indeed escape from a funnel will become more difficult with increasing size because the
interfunnel barriers become larger.

Therefore, during cluster growth, except at high temperatures and extremely long time scales, 
the smaller clusters serves as templates for growth and
only the optimization of the position of additional loosely-bound molecules is likely to occur.
If icosahedral structures are preferred at the last size for which equilibrium is possible, then 
larger icosahedra will result. Thus, the disconnectivity graphs give additional
support to the idea that icosahedra are observed experimentally because of kinetic trapping 
at low temperature.

\begin{figure}
\begin{center}
\includegraphics[width=8.2cm]{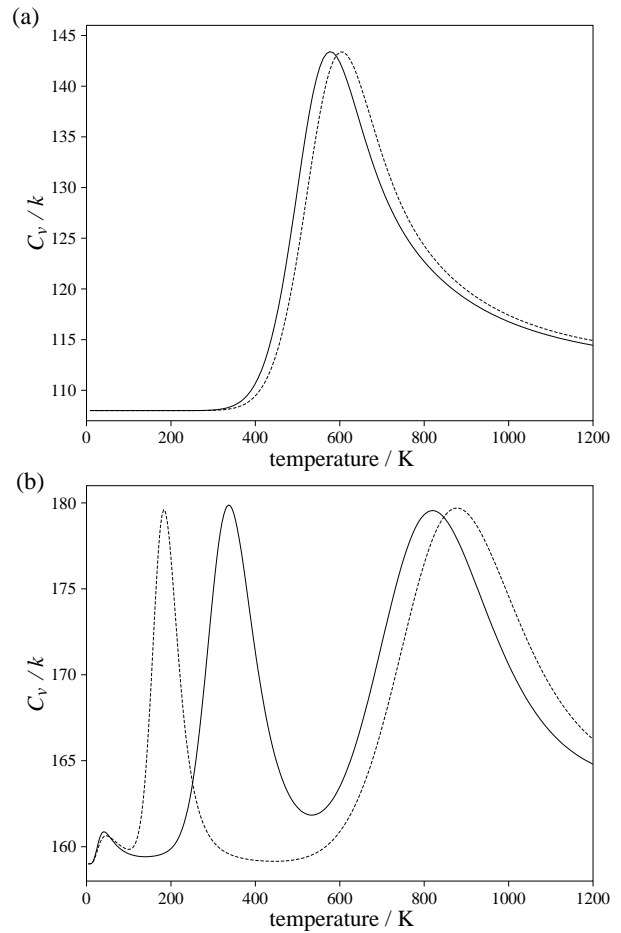}
\caption{\label{fig:Cv} Heat capacity curves for (a) (C$_{60}$)$_{38}$ and (C$_{60}$)$_{55}$ 
obtained from our samples of minima using the harmonic superposition method. The
solid lines are for the full potential and the dashed lines are for the two-body potential.}
\end{center}
\end{figure}

To examine the low-temperature thermodynamics of the two clusters in question we use the
harmonic superposition method.\cite{Wales93a,WalesDMMW00} 
In this approach the partition function is constructed by 
summing the partition functions of the individual minima on the PES. In
the harmonic approximation, we obtain
\begin{equation}
Z=\sum_i {n_i \exp\left(-\beta E_i/kT\right)\over \left(\beta h \overline{\nu}_i\right)^{3N-6}},
\end{equation}
where $E_i$ is the energy of minimum $i$, $\overline{\nu}_i$ is the geometric mean frequency
and $n_i$ is the number of permutational isomers ($2N!/h_i$, where $h_i$ is the order of the point group).

This approach is unable to reproduce high temperature behaviour such as melting in the present cases,
because no effort has been made to compensate for the incompleteness of our samples of minima, 
thus leading to an underestimate of the configurational entropy of the high-energy states. 
It is possible to overcome this limitation by using information obtained from 
an ergodic simulation,\cite{Wales93a,WalesDMMW00} but this is not attempted because ergodicity is 
hard to achieve for these clusters and because we are more interested in solid-solid 
transitions, for which we only need to characterize the low-energy regions of the PES. 
Instead we used parallel-tempering to locate the point at which the cluster melts or boils.
It is also possible to introduce anharmonicity into this scheme,\cite{Doye95a,Calvo01e} however
this additional level of complexity is unnecessary for the insights we are seeking to achieve here.

The advantage of the superposition method is the ease with which the low-temperature thermodynamics
can be obtained even when, as with (C$_{60}$)$_N$ clusters, 
there are large interfunnel energy barriers present.
In fact it is not clear if any other method would be able to probe this regime, 
because of the difficulty of achieving ergodicity.
Even techniques based on parallel tempering,\cite{Swendsen86,Marinari92} which are the only direct simulation
methods that have been shown to succeed for challenging cases involving low-temperature solid-solid 
transitions in clusters,\cite{Neirotti00,Calvo01a} failed for (C$_{60}$)$_{55}$.
The success of these methods relies on the coupling of the low-temperature runs 
to ergodic high-temperature runs involving molten clusters. However, because of the narrowness of any range
of stability for the liquid-like state of clusters of C$_{60}$ molecules\cite{Gallego99,Calvo01c} 
this is very hard to achieve.

Heat capacity curves for the two clusters are presented in Figure \ref{fig:Cv}.
These were calculated using the same samples of minima as for the disconnectivity graphs.
It was also possible to reoptimize all the minima for the full potential.
Both clusters show some evidence of solid-solid transitions. 

For (C$_{60}$)$_{55}$ the energy gap between the icosahedral minima and the global minimum 
is relatively small and so a transition to the Mackay icosahedron occurs significantly below the melting temperature.
For the full potential this transition occurs at 377$\,$K and, consistent with the reduction in the energy gap, 
at 184$\,$K when the three-body term is neglected.
Interestingly, there is also a small peak at $\sim$45$\,$K, which corresponds to a redistribution of the 
equilibrium occupation probability amongst the five lowest-energy decahedral isomers. These
minima are almost isoenergetic and correspond to the five different positions that 
the four-coordinate surface molecules can occupy. 
The high-temperature peak corresponds to the beginnings of the melting/boiling transition,
which as expected is underestimated by the harmonic superposition method. 
Parallel tempering indicates that the actual peak for this transition is much larger 
and occurs at $\sim$950$\,$K, although the result is somewhat
dependent on the size of the constraining box in which the cluster is placed. 

For (C$_{60}$)$_{38}$ the calculated heat capacity curve is much simpler and only has a single peak.
This corresponds to the transition out of the truncated octahedron, first roughly equally 
into the low-energy decahedral minima and lowest-energy icosahedral minimum, and then at slightly
higher temperature into the higher-energy states. As these transitions are so close together in
temperature there is only a single peak in the heat capacity, which is centred on the initial 
transition because of the underestimation of the latent heat of melting by the present approach. 
Parallel tempering simulations suggest that the preliminary solid-solid transition 
gives rise to a shoulder in the melting peak, which occurs at $\sim$900$\,$K.

Although we have only considered two sizes in detail, the large temperature window for which the 
Mackay icosahedron is most stable indicates that solid-solid transitions 
are likely to be common for other sizes.
The results confirm what has been emphasized elsewhere, namely that crossover sizes at which 
the dominant structure changes depend on temperature.\cite{Doye01b}
So, although none of the clusters above $N$=15 have icosahedral global minima, 
icosahedra can still be thermodynamically most stable for some range of temperature above this size.
For example, similar calculations indicate that for (C$_{60}$)$_{19}$ icosahedral structures 
are favoured above 333$\,$K.
However, these results cannot explain the experimental 
observation of icosahedra, because they are seen at low not high temperature.

As for the solid-solid transitions that have been investigated in detail
for Lennard-Jones clusters,\cite{Doye01b,Doye01f} the above transitions are driven
by the larger vibrational entropy of the icosahedral structures. For $N$=38
$\overline{\nu}_{\rm icos}$:$\overline{\nu}_{\rm deca}$:$\overline{\nu}_{\rm cp}$=
0.920:0.977:1
and for $N$=55 
0.909:1:0.999,
where we have used the values for the lowest-energy minimum 
of each structural type. These differences are much larger than for Lennard-Jones clusters,
where, for example,
$\overline{\nu}_{\rm icos}$:$\overline{\nu}_{\rm deca}$:$\overline{\nu}_{\rm cp}$=1:1.000:0.990 for LJ$_{55}$
and the systematic differences between compact sequences of structures are no larger than 2\% 
(Mackay icosahedra having the lowest frequencies).\cite{Doye01f}

\begin{figure}
\begin{center}
\includegraphics[width=8.2cm]{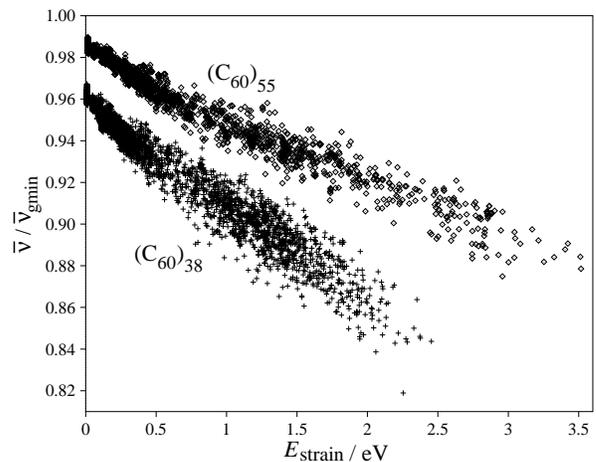}
\caption{\label{fig:strain} The dependence of $\overline{\nu}$ on $E_{\rm strain}$ for 
(C$_{60}$)$_{38}$ minima with $n_{\rm nn}$=139 (crosses) and 
(C$_{60}$)$_{55}$ minima with $n_{\rm nn}$=217 (diamonds). 
These samples contain 8408 and 7337 minima, respectively, and correspond to the largest subsets of the
complete samples that have the same value of $n_{\rm nn}$. The mean frequency is measured with respect to
that for the global minimum.}
\end{center}
\end{figure}

$\overline\nu$ is a measure of how the energy increases as the structure is distorted, and
so we can gain insight into the origins of the dependence of $\overline{\nu}$ on structure if we examine
the different contributions to the energy. For a pair potential
\begin{equation}
E=-n_{\rm nn}\epsilon + E_{\rm strain}+ E_{\rm nnn}, 
\end{equation}
where $n_{\rm nn}$ is the number of nearest neighbours 
(defined using a distance criterion $r_0$), the strain energy, the energetic penalty for 
distances deviating from the equilibrium pair distance, is given by  
\begin{equation}
E_{\rm strain}=\sum_{i<j,r_{ij}<r_0} V(r_{ij})+\epsilon 
\end{equation}
and the energy of non-nearest neighbours by  
\begin{equation}
E_{\rm nnn}=\sum_{i<j,r_{ij}>r_0} V(r_{ij}).
\end{equation}

As $E_{\rm nnn}$ is small for a short-ranged potential and does not vary strongly with distance
the main contributions to $\overline\nu$ come from nearest neighbours. Therefore, it is clear that
the more nearest neighbours a structure has the greater the effect on the energy. Thus, compact
structures with few low-coordinate surface molecules have larger vibrational frequencies. This is the
reason that $\overline\nu$ generally decreases as the energy of the minima increases. 

The strain energy also has a significant effect on the vibrational frequencies. 
For an unstrained structure all nearest neighbours lie at the equilibrium separation, 
and so any distortion of the structure leads to an increase 
in all the individual nearest-neighbour pair energies.
For a strained structure, there is a dispersion of nearest-neighbour distances and 
so a distortion is likely to cause some of them to 
deviate further from the equilibrium distance, but others to become closer. 
This compensation leads to a smaller rise in the energy, and hence to smaller vibrational frequencies. 
When the effect is isolated by only comparing minima with the same number of nearest neighbours, 
this trend is clear (Figure \ref{fig:strain}). 

As the curvature of the pair potential increases, the effect of strain becomes larger.
We can therefore explain both the much lower frequencies for the icosahedral structures 
in our two examples and why this effect is more dramatic than for Lennard-Jones clusters.
Therefore, as the range of the potential is decreased, the increasing energy gap between icosahedra
and the lowest-energy structures is compensated by a lower relative mean vibrational frequency
and so it is still possible that icosahedral structures are stable for some range of temperature,
even when the potential is relatively short-ranged.

The differences in $\overline{\nu}$ between close-packed and decahedral structures are smaller, because
of the smaller differences in the strain energy. Indeed, for (C$_{60}$)$_{55}$ the lowest-energy
decahedral minima actually has a higher $\overline{\nu}$ because the effect of a larger $E_{\rm strain}$ 
is more than compensated by the effect of a greater $n_{\rm nn}$.

As Leary tetrahedra have a strain energy intermediate between icosahedra and decahedra, one would expect 
that their mean vibrational frequency is lower than for decahedral and close-packed clusters.
Indeed, this is the case. For example, for (C$_{60}$)$_{98}$ 
$\overline{\nu}_{\rm icos}$:$\overline{\nu}_{\rm Leary}$:$\overline{\nu}_{\rm deca}$:$\overline{\nu}_{\rm cp}$=
0.925:0.972:0.997:1.
Although the Leary tetrahedron does develop a small equilibrium population near to the melting point in 
the current model, there is no clear transition at which the Leary tetrahedron becomes the dominant structure.

\section{Conclusions}
\label{sect:conclusion}
Putative global minima for the C$_{60}$ intermolecular potential of Pacheco and Prates-Ramalho
have been located for all clusters containing up to 105 molecules. For $N\le 15$ the structures follow an
icosahedral growth sequence, but above this size the global minima are either decahedral
or close-packed, with close-packed clusters becoming more common as the size increases. 
This progression to structures with lower strain energy as the size increases
is expected, but is more rapid than for most atomic clusters, because the PPR potential 
is much more short-ranged than typical interatomic potentials.

The correspondence between many of the particularly low-energy PPR clusters and the
high temperature experimental magic numbers adds further support to the interpretation
that these sizes have particularly low free energies because they are particularly
low in potential energy. This interpretation also implies that the icosahedral magic numbers 
seen at lower temperature are kinetic in origin. 
The large interfunnel barriers evident from the disconnectivity graphs
of (C$_{60}$)$_{38}$ and (C$_{60}$)$_{55}$ add further support to this hypothesis
by indicating the difficulty of major structural transformations in a growing cluster.
Moreover, preliminary results from growth simulations confirm this scenario.\cite{growth} 

However, there are still discrepancies between the high temperature magic numbers and
the low-energy model C$_{60}$ clusters. Firstly, the magic number at $N$=19, which 
persists up to high temperature, is probably due to the double icosahedron. 
This perhaps suggests that the PPR potential is still effectively too short-ranged,
but an alternative explanation is the thermal stabilization of this structure at the
experimental temperature. 

Secondly, a particular preference for structures that are based 
on the Leary tetrahedron is not reproduced. 
It is possible that these structures are only lowest in free energy at high temperature,
but this seems unlikely for two reasons. 
Firstly, as the entropy is not so strongly size-dependent as the energy, 
magic numbers associated with structures that are entropically preferred are likely 
to be less pronounced. However, in the present case the magic numbers associated
with the structures based upon the Leary tetrahedron are the most prominent. 
Secondly, no such transition is seen for (C$_{60}$)$_{98}$.
Hence, it is more likely that the discrepancy is due to the potential, and
again suggests that the PPR potential is effectively too short-ranged. For 
the Morse potential the 98-atom Leary tetrahedron is lowest in energy for $6.91<\rho<9.45$.
However, the situation is somewhat more complex.
Only at a few sizes are structures based on the Leary tetrahedron actually 
the global minimum of the Morse potential.
Therefore, the observed preference cannot simply be explained by the effective range. 
It is most likely to be related to the orientational degrees of freedom that
are neglected in a single-site potential, such as the PPR model. 

The thermodynamics calculations also indicate further discrepancies between 
experiment and the PPR model. We saw in Section \ref{sect:3855} that icosahedral structures are 
thermodynamically favoured at high temperature for certain sizes, because of their 
larger vibrational entropy. However, except perhaps for $N$=19, there is no experimental 
support for this scenario.

Although all-atom potentials have been employed for modelling clusters of (C$_{60}$) 
molecules,\cite{Rey97,Doye97c,Garcia97} the double icosahedron was found not to be 
lowest in energy for $N$=19 and the potentials were not applied in the size range relevant to 
structures based on the Leary tetrahedron. Furthermore, these types of all-atom 
potential are unable to reproduce the low-temperature properties of 
bulk C$_{60}$---the molecules have the incorrect orientation in the low-temperature 
crystal.\cite{Cheng92c}
Although, at the temperatures relevant to experiment, the molecules are expected to be able to
rotate freely,\cite{Deleuze99} it could well be that the energetic preferences 
for structures in which the molecules can have the preferred orientations 
will persist up to higher temperature.
Therefore, to reproduce the preference for Leary tetrahedra it is likely that 
a potential that can give the correct orientations in the crystal is required.
There are a number of such potentials,\cite{Lu92,Sprik92,Gamba98} however
this orientational preference has sometimes been achieved through the introduction 
of unrealistic electrostatic properties for the C$_{60}$ molecule.\cite{Yildirim93}
Still, it would be interesting to know if these potentials do favour
Leary tetrahedra. However, this would be an extremely challenging
task computationally, both because of the complexities of the potentials and
the huge number of possible orientational isomers for such large clusters.

\acknowledgements
JPKD is grateful to Emmanuel College, Cambridge for financial support.

\begin{table*}
\caption{\label{table:gmin}Energies and point groups for the putative global minima 
for (C$_{60}$)$_N$ clusters modelled by the PPR potential. 
The energies of these minima when reoptimized
without the inclusion of the three-body term ($E_2$)
are also given.
Structural assignments are given for all clusters with $N\ge 13$, where i stands for 
icosahedral, d for decahedral (d$^+$ have an overlayer on the $\{111\}$ faces), f for 
fcc and c for close-packed (but not fcc). Fcc and close-packed structures can be unstrained.}
\begin{ruledtabular}
\begin{tabular}{ccccccccccccccccc}
$N$ & PG & S & $E$/eV & $E_2$/eV & & $N$ & PG & S & $E$/eV & $E_2$ & & $N$ & PG & S & $E$/eV & $E_2$/eV \\
\hline
   3 & $D_{3h}$  & &    -0.793763 &    -0.800181 & &   38 & $O_h$     & f &   -38.726684 &   -40.164289 & &   73 & $C_s$     & d &   -82.892601 &   -86.232181 \\ 
   4 & $T_d$     & &    -1.574722 &    -1.600362 & &   39 & $C_{4v}$  & f &   -39.803068 &   -41.276819 & &   74 & $C_{5v}$  & d &   -84.275939 &   -87.683469 \\ 
   5 & $D_{3h}$  & &    -2.360685 &    -2.406624 & &   40 & $C_{2v}$  & f &   -40.879585 &   -42.389454 & &   75 & $D_{5h}$  & d &   -85.854200 &   -89.335316 \\ 
   6 & $O_h$     & &    -3.179823 &    -3.251103 & &   41 & $C_{2v}$  & d &   -42.050754 &   -43.620062 & &   76 & $C_s$     & d &   -86.943730 &   -90.469387 \\ 
   7 & $D_{5h}$  & &    -4.174831 &    -4.277606 & &   42 & $C_s$     & d &   -43.134640 &   -44.743772 & &   77 & $C_{2v}$  & d &   -88.287170 &   -91.870873 \\ 
   8 & $C_s$     & &    -4.971222 &    -5.095539 & &   43 & $C_{2v}$  & d &   -44.465028 &   -46.115661 & &   78 & $C_1$     & d &   -89.376306 &   -93.004550 \\ 
   9 & $C_{2v}$  & &    -5.979744 &    -6.139683 & &   44 & $C_s$     & d &   -45.548888 &   -47.250124 & &   79 & $D_{3h}$  & c &   -90.750950 &   -94.414956 \\ 
  10 & $C_{3v}$  & &    -6.941044 &    -7.136890 & &   45 & $C_{2v}$  & d &   -46.879111 &   -48.632640 & &   80 & $C_s$     & c &   -91.847206 &   -95.555790 \\ 
  11 & $C_{2v}$  & &    -7.890403 &    -8.122285 & &   46 & $C_s$     & c &   -47.964493 &   -49.771293 & &   81 & $C_{2v}$  & c &   -93.195399 &   -96.961852 \\ 
  12 & $C_{5v}$  & &    -8.910974 &    -9.191726 & &   47 & $C_{2v}$  & d &   -49.293304 &   -51.138897 & &   82 & $C_{2v}$  & d &   -94.475899 &   -98.305791 \\ 
  13 & $I_h$     & i &   -10.203987 &   -10.556666 & &   48 & $C_{2v}$  & d$^+$ &   -50.835900 &   -52.786358 & &   83 & $C_{2v}$  & c &   -95.639897 &   -99.508789 \\ 
  14 & $C_{3v}$  & i &   -11.010918 &   -11.384715 & &   49 & $C_s$     & d$^+$ &   -51.926724 &   -53.919060 & &   84 & $C_{2v}$  & d &   -96.907100 &  -100.839577 \\ 
  15 & $C_{2v}$  & i &   -12.028948 &   -12.436957 & &   50 & $D_{3h}$  & c &   -53.354529 &   -55.400213 & &   85 & $C_{3v}$  & c &   -98.138499 &  -102.126340 \\ 
  16 & $C_{2v}$  & d &   -13.017348 &   -13.429036 & &   51 & $C_s$     & c &   -54.449934 &   -56.540311 & &   86 & $C_{3v}$  & c &   -99.728365 &  -103.787568 \\ 
  17 & $C_{2v}$  & d &   -14.088380 &   -14.538481 & &   52 & $C_{2v}$  & c &   -55.797330 &   -57.945711 & &   87 & $C_s$     & c &  -100.867536 &  -104.983413 \\ 
  18 & $D_{5h}$  & d &   -15.163644 &   -15.653781 & &   53 & $C_s$     & c &   -56.892907 &   -59.085964 & &   88 & $C_s$     & c &  -102.173549 &  -106.335126 \\ 
  19 & $C_{2v}$  & d &   -16.218799 &   -16.743185 & &   54 & $C_{2v}$  & d &   -58.306782 &   -60.570092 & &   89 & $C_s$     & d$^+$ &  -103.461268 &  -107.695215 \\ 
  20 & $C_{2v}$  & d &   -17.274543 &   -17.832640 & &   55 & $C_{2v}$  & d &   -59.408665 &   -61.716335 & &   90 & $C_s$     & d$^+$ &  -104.807341 &  -109.095407 \\ 
  21 & $C_s$     & d$^+$ &   -18.353074 &   -18.949300 & &   56 & $C_s$     & d &   -60.752438 &   -63.116143 & &   91 & $C_{3v}$  & c &  -106.302975 &  -110.678677 \\ 
  22 & $C_1$     & d$^+$ &   -19.425310 &   -20.059762 & &   57 & $C_{2v}$  & d &   -62.098302 &   -64.517510 & &   92 & $C_s$     & c &  -107.397152 &  -111.806846 \\ 
  23 & $C_s$     & d$^+$ &   -20.686563 &   -21.386861 & &   58 & $D_{3h}$  & d$^+$ &   -63.386632 &   -65.901565 & &   93 & $C_s$     & c &  -108.747951 &  -113.216329 \\ 
  24 & $C_1$     & d$^+$ &   -21.759634 &   -22.497858 & &   59 & $T_d$     & c &   -64.769800 &   -67.323349 & &   94 & $C_s$     & c &  -110.100130 &  -114.626428 \\ 
  25 & $C_{3v}$  & d$^+$ &   -22.957151 &   -23.760958 & &   60 & $C_s$     & d &   -65.887129 &   -68.480038 & &   95 & $C_{2v}$  & d &  -111.653413 &  -116.285755 \\ 
  26 & $C_{2v}$  & d$^+$ &   -24.141029 &   -24.968567 & &   61 & $C_{3v}$  & f &   -67.169082 &   -69.798058 & &   96 & $C_s$     & d &  -112.771588 &  -117.460346 \\ 
  27 & $C_s$     & d$^+$ &   -25.218788 &   -26.085977 & &   62 & $C_1$     & d &   -68.330220 &   -71.021579 & &   97 & $C_1$     & d &  -114.123353 &  -118.862596 \\ 
  28 & $C_s$     & d$^+$ &   -26.403185 &   -27.334120 & &   63 & $C_s$     & d &   -69.702863 &   -72.458970 & &   98 & $C_s$     & c &  -115.518307 &  -120.295415 \\ 
  29 & $D_{5h}$  & d &   -27.542310 &   -28.503488 & &   64 & $C_{2v}$  & d &   -71.278073 &   -74.107991 & &   99 & $C_s$     & d &  -117.079750 &  -121.959509 \\ 
  30 & $C_{2v}$  & d$^+$ &   -28.649061 &   -29.682832 & &   65 & $C_{2v}$  & d &   -72.382930 &   -75.268852 & &  100 & $T_d$     & c &  -118.491490 &  -123.409546 \\ 
  31 & $C_{2v}$  & d &   -29.963558 &   -31.028821 & &   66 & $C_s$     & d &   -73.720048 &   -76.664850 & &  101 & $D_{5h}$  & d &  -120.052347 &  -125.073354 \\ 
  32 & $C_1$     & d &   -31.047706 &   -32.157893 & &   67 & $C_{2v}$  & d &   -75.059516 &   -78.062777 & &  102 & $C_{2v}$  & d &  -121.170243 &  -126.247600 \\ 
  33 & $C_{2v}$  & d &   -32.386698 &   -33.555532 & &   68 & $C_1$     & d &   -76.157881 &   -79.200432 & &  103 & $C_s$     & d &  -122.516174 &  -127.652314 \\ 
  34 & $C_1$     & d &   -33.469737 &   -34.683456 & &   69 & $C_1$     & d &   -77.504125 &   -80.605732 & &  104 & $C_{2v}$  & d &  -123.864106 &  -129.058381 \\ 
  35 & $C_{2v}$  & d &   -34.807885 &   -36.080228 & &   70 & $C_s$     & d &   -78.872607 &   -82.039199 & &  105 & $C_s$     & d &  -124.981491 &  -130.232109 \\ 
  36 & $C_1$     & d &   -35.889548 &   -37.206754 & &   71 & $C_{2v}$  & d &   -80.450302 &   -83.690481 & \\
  37 & $C_{2v}$  & d &   -37.226573 &   -38.602386 & &   72 & $C_1$     & d &   -81.545223 &   -84.828030 & \\
\end{tabular}
\end{ruledtabular}
\end{table*}


\begin{thebibliography}{10}

\bibitem{Hagen93}
M.~H.~J. Hagen {\it et~al.}, Nature {\bf 365},  425  (1993).

\bibitem{Broughton97}
J.~Q. Broughton, J.~V. Lill, and J.~K. Johnson, Phys. Rev. B {\bf 55},  2808
  (1997).

\bibitem{ACheng}
A. Cheng, M.~L. Klein, and C. Caccamo, Phys. Rev. Lett. {\bf 71},  1200
  (1993).

\bibitem{Caccamo97}
C. Caccamo, D. Costa, and A. Fucile, J. Chem. Phys. {\bf 106},  255  (1997).

\bibitem{Hasegawa99}
M. Hasegawa and K. Ohno, J. Chem. Phys. {\bf 111},  5955  (1999).

\bibitem{Ferreira00}
A.~L.~C. Ferreira, J.~M. Pacheco, and J.~P. Prates-Ramalho, J. Chem. Phys. {\bf
  113},  738  (2000).

\bibitem{Hasegawa00}
M. Hasegawa and K. Ohno, J. Chem. Phys. {\bf 113},  4315  (2000).

\bibitem{Martin93}
T.~P. Martin, U. N\"aher, H. Schaber, and U. Zimmermann, Phys. Rev. Lett. {\bf
  70},  3079  (1993).

\bibitem{Mackay}
A.~L. Mackay, Acta Cryst. {\bf 15},  916  (1962).

\bibitem{Hansen96}
K. Hansen, H. Hohmann, R. M\"{u}ller, and E.~E.~B. Campbell, J. Chem. Phys.
  {\bf 105},  6088  (1996).

\bibitem{Hansen97}
K. Hansen, H. Hohmann, R. M\"{u}ller, and E.~E.~B. Campbell, Z. Phys. D {\bf
  40},  361  (1997).

\bibitem{Girifalco}
L.~A. Girifalco, J. Phys. Chem. {\bf 96},  858  (1992).

\bibitem{Rey94}
C. Rey, L.~J. Gallego, and J.~A. Alonso, Phys. Rev. B {\bf 49},  8491  (1994).

\bibitem{Wales94a}
D.~J. Wales, J. Chem. Soc., Faraday Trans. {\bf 90},  1061  (1994).

\bibitem{Doye96d}
J.~P.~K. Doye and D.~J. Wales, Chem. Phys. Lett. {\bf 262},  167  (1996).

\bibitem{Martin90}
T.~P. Martin, T. Bergmann, H. G\"ohlich, and T. Lange, Chem. Phys. Lett. {\bf
  172},  209  (1990).

\bibitem{Doye95c}
J.~P.~K. Doye, D.~J. Wales, and R.~S. Berry, J. Chem. Phys. {\bf 103},  4234
  (1995).

\bibitem{Doye96a}
J.~P.~K. Doye and D.~J. Wales, Science {\bf 271},  484  (1996).

\bibitem{Doye96b}
J.~P.~K. Doye and D.~J. Wales, J. Phys. B {\bf 29},  4859  (1996).

\bibitem{Rey97}
C. Rey, J. Garcia-Rodeja, and L.~J. Gallego, Z. Phys. D {\bf 40},  395  (1997).

\bibitem{Doye97c}
J.~P.~K. Doye, A. Dullweber, and D.~J. Wales, Chem. Phys. Lett. {\bf 269},  408
   (1997).

\bibitem{Garcia97}
J. Garcia-Rodeja, C. Rey, and L.~J. Gallego, Phys. Rev. B {\bf 56},  6466
  (1997).

\bibitem{Shvartsburg96}
A.~A. Shvartsburg and M.~F. Jarrold, Chem. Phys. Lett. {\bf 86},  261  (1996).

\bibitem{Branz00}
W. Branz, N. Malinowski, H. Schaber, and T.~P. Martin, Chem. Phys. Lett. {\bf
  328},  245  (2000).

\bibitem{timescale}
These temperatures are related to the time the clusters spend in the heating
  cell, which was approximately 0.5$\,$ms.\cite{Branz00} If the heating time in
  the experiment is increased the temperatures for the observation of the
  icosahedral and non-icosahedral structures decrease somewhat. For example,
  for a heating time of 1$\,$ms the decreases are 15-20$\,$K.

\bibitem{Leary99}
R.~H. Leary and J.~P.~K. Doye, Phys. Rev. E {\bf 60},  R6320  (1999).

\bibitem{Doye98a}
J.~P.~K. Doye and D.~J. Wales, Phys. Rev. Lett. {\bf 80},  1357  (1998).

\bibitem{Cleveland98}
C.~L. Cleveland, W.~D. Luedtke, and U. Landman, Phys. Rev. Lett. {\bf 81},
  2036  (1998).

\bibitem{Cleveland99}
C.~L. Cleveland, W.~D. Luedtke, and U. Landman, Phys. Rev. B {\bf 60},  5065
  (1999).

\bibitem{Doye99c}
J.~P.~K. Doye, M.~A. Miller, and D.~J. Wales, J. Chem. Phys. {\bf 110},  6896
  (1999).

\bibitem{Neirotti00}
J.~P. Neirotti, F. Calvo, D.~L. Freeman, and J.~D. Doll, J. Chem. Phys. {\bf
  112},  10340  (2000).

\bibitem{Doye01b}
J.~P.~K. Doye and F. Calvo, Phys. Rev. Lett. {\bf 86},  3570  (2001).

\bibitem{Calvo01d}
F. Calvo, J.~P.~K. Doye, and D.~J. Wales, Phys. Rev. Lett.  submitted  (2001).

\bibitem{Zhang00}
W. Zhang, L. Liu, J. Zhuang, and Y. Li, Phys. Rev. B {\bf 62},  8276  (2000).

\bibitem{Baletto01}
F. Baletto, C. Mottet, and R. Ferrando, Phys. Rev. B {\bf 63},  155408  (2001).

\bibitem{Doye99f}
J.~P.~K. Doye, M.~A. Miller, and D.~J. Wales, J. Chem. Phys. {\bf 111},  8417
  (1999).

\bibitem{Northby87}
J.~A. Northby, J. Chem. Phys. {\bf 87},  6166  (1987).

\bibitem{Wales94b}
D.~J. Wales, J. Chem. Phys. {\bf 101},  3750  (1994).

\bibitem{Miller99a}
M.~A. Miller, J.~P.~K. Doye, and D.~J. Wales, J. Chem. Phys. {\bf 110},  328
  (1999).

\bibitem{Miller99b}
M.~A. Miller, J.~P.~K. Doye, and D.~J. Wales, Phys. Rev. E {\bf 60},  3701
  (1999).

\bibitem{Gallego99}
L.~J. Gallego, J. Garcia-Rodeja, M.~M.~G. Alemany, and C. Rey, Phys. Rev. Lett.
  {\bf 83},  5258  (1999).

\bibitem{Pacheco97}
J.~M. Pacheco and J.~P. Prates-Ramalho, Phys. Rev. Lett. {\bf 79},  3873
  (1997).

\bibitem{Calvo01c}
F. Calvo, J. Phys. Chem. B {\bf 105},  2183  (2001).

\bibitem{Luo99}
Y.-H. Luo, J. Zhao, S. Qiu, and G. Wang, Phys. Rev. B {\bf 59},  14903  (1999).

\bibitem{Axilrod}
B.~M. Axilrod and E. Teller, J. Chem. Phys. {\bf 11},  299  (1943).

\bibitem{WalesU94}
D.~J. Wales and J. Uppenbrink, Phys. Rev. B {\bf 50},  12342  (1994).

\bibitem{Hagen94}
M.~H.~J. Hagen and D. Frenkel, J. Chem. Phys. {\bf 101},  4093  (1994).

\bibitem{Doye97d}
J.~P.~K. Doye and D.~J. Wales, J. Chem. Soc., Faraday Trans. {\bf 93},  4233
  (1997).

\bibitem{Wales90a}
D.~J. Wales, J. Chem. Soc., Faraday Trans. {\bf 86},  3505  (1990).

\bibitem{Up92}
J. Uppenbrink and D.~J. Wales, J. Chem. Phys. {\bf 96},  8520  (1992).

\bibitem{WalesD97}
D.~J. Wales and J.~P.~K. Doye, J. Phys. Chem. A {\bf 101},  5111  (1997).

\bibitem{Li87a}
Z. Li and H.~A. Scheraga, Proc. Natl. Acad. Sci. USA {\bf 84},  6611  (1987).

\bibitem{WalesS99}
D.~J. Wales and H.~A. Scheraga, Science {\bf 285},  1368  (1999).

\bibitem{Doye98e}
J.~P.~K. Doye, D.~J. Wales, and M.~A. Miller, J. Chem. Phys. {\bf 109},  8143
  (1998).

\bibitem{Doye01c}
J.~P.~K. Doye,  in {\em Global Optimization---Selected Case Studies}, edited by
  J.~D. Pinter (Kluwer Academic, Dordrecht, 2001).

\bibitem{Hartke96}
B. Hartke, Chem. Phys. Lett. {\bf 258},  144  (1996).

\bibitem{Doye00c}
J.~P.~K. Doye, Phys. Rev. E {\bf 62},  8753  (2000).

\bibitem{Cerjan}
C.~J. Cerjan and W.~H. Miller, J. Chem. Phys. {\bf 75},  2800  (1981).

\bibitem{WalesDMMW00}
D.~J. Wales {\it et~al.}, Adv. Chem. Phys. {\bf 115},  1  (2000).

\bibitem{Web}
D. J. Wales, J. P. K. Doye, A. Dullweber, M. P. Hodges, F. Y. Naumkin and F.
  Calvo, The Cambridge Cluster Database, URL
  http://www-wales.ch.cam.ac.uk/CCD.html.

\bibitem{Baletto00}
F. Baletto, C. Mottet, and R. Ferrando, Phys. Rev. Lett. {\bf 84},  5544
  (2000).

\bibitem{StillD90a}
F.~H. Stillinger and D.~K. Stillinger, J. Chem. Phys. {\bf 93},  6013  (1990).

\bibitem{Kunz93}
R.~E. Kunz and R.~S. Berry, Phys. Rev. Lett. {\bf 71},  3987  (1993).

\bibitem{Becker97}
O.~M. Becker and M. Karplus, J. Chem. Phys. {\bf 106},  1495  (1997).

\bibitem{WalesMW98}
D.~J. Wales, M.~A. Miller, and T.~R. Walsh, Nature {\bf 394},  758  (1998).

\bibitem{Leopold}
P.~E. Leopold, M. Montal, and J.~N. Onuchic, Proc. Natl. Acad. Sci. USA {\bf
  89},  8271  (1992).

\bibitem{Bryngelson95}
J.~D. Bryngelson, J.~N. Onuchic, N.~D. Socci, and P.~G. Wolynes, Proteins:
  Structure, Function and Genetics {\bf 21},  167  (1995).

\bibitem{NelsonS}
D.~R. Nelson and F. Spaepen, Solid State Phys. {\bf 42},  1  (1989).

\bibitem{Wales93a}
D.~J. Wales, Mol. Phys. {\bf 78},  151  (1993).

\bibitem{Doye95a}
J.~P.~K. Doye and D.~J. Wales, J. Chem. Phys. {\bf 102},  9659  (1995).

\bibitem{Calvo01e}
F. Calvo, J.~P.~K. Doye, and D.~J. Wales, J. Chem. Phys.  submitted  (2001).

\bibitem{Marinari92}
E. Marinari and G. Parisi, Europhys. Lett. {\bf 19},  451  (1992).

\bibitem{Swendsen86}
R.~H. Swendsen and J.-S. Wang, Phys. Rev. Lett. {\bf 57},  2607  (1986).

\bibitem{Calvo01a}
F. Calvo and J.~P.~K. Doye, Phys. Rev. E {\bf 63},  010902(R)  (2001).

\bibitem{Doye01f}
J.~P.~K. Doye and F. Calvo, J. Chem. Phys.  to be submitted  .

\bibitem{growth}
F. Baletto, J. P. K. Doye and R. Ferrando, unpublished.

\bibitem{Cheng92c}
A. Cheng and M.~L. Klein, Phys. Rev. B {\bf 45},  1889  (1992).

\bibitem{Deleuze99}
M.~S. Deleuze and F. Zerbetto, J. Am. Chem. Soc. {\bf 121},  5281  (1999).

\bibitem{Gamba98}
Z. Gamba, Phys. Rev. B {\bf 57},  1402  (1998).

\bibitem{Lu92}
J.~P. Lu, X.-P. Li, and R.~M. Martin, Phys. Rev. Lett. {\bf 68},  1551  (1992).

\bibitem{Sprik92}
M. Sprik, A. Cheng, and M.~L. Klein, J. Phys. Chem. {\bf 96},  2027  (1992).

\bibitem{Yildirim93}
T. Yildirim, A.~B. Harris, S.~C. Erwin, and M.~R. Pederson, Phys. Rev. B {\bf
  48},  1888  (1993).

\end{thebibliography}
\end{document}